\documentclass[multphys,vecphys]{svmult}
\usepackage{epsfig,psfrag,amsmath,amssymb,float}
\usepackage{makeidx}     
\usepackage{graphicx}    

\usepackage{multicol}    

\makeindex             

\begin{document}

\title*{Simulations of pure and doped low-dimensional spin-1/2 gapped systems}
\author{Nicolas Laflorencie and Didier Poilblanc}
\institute{Laboratoire de Physique Th\'eorique, CNRS-UMR5152
Universit\'e Paul Sabatier, F-31062 Toulouse, France
\texttt{didier.poilblanc@irsamc.ups-tlse.fr,~nicolas.laflorencie@irsamc.ups-tlse.fr}}
%
%
\maketitle

\section{Introduction}
\label{sec:1}

Many systems of Condensed Matter consist of fermions (electrons)
moving on a lattice and experiencing strong repulsive 
interactions~\cite{Fulde}.
In such cases, the traditional perturbative methods to treat the electronic 
correlations often break down. In a pioneering work, Bonner and Fisher 
\cite{Bonner_Fisher} revealed the exact diagonalisation (ED) method 
as a powerful tool to study the properties of one dimensional (1D)
spin chains. Later on, it was extended to investigate
two dimensional (2D) localised spin systems \cite{Oitmaa_Betts}.
This work initiated a more extensive use of the method 
to investigate a wide variety of different systems such as strongly 
correlated lattice electrons 
(Hubbard-like models), mesoscopic systems~\cite{mesoscopic},
electron-phonon models, etc.... 

The success of this method first comes from the rapidly growing power
of supercomputers which are being equipped with faster and faster processors 
and larger and larger memories, disk space and storage facilities.  
In addition, ED are clearly {\it unbiased} methods as we shall discuss later
on in the course of this Chapter. The systematic errors can be, in most cases,
easily estimated and, hence, this method is a very controlled one. 
Of course, it has its limitations (which we shall also discuss later) but,
clearly, the efficiency of this technique will steadily increase in the future
as the power of supercomputers booms up. 

The following Chapter will be dedicated mostly to the numerical technique based on the Lanczos algorithm. However, we shall focus on a specific area of strongly correlated models, namely low-dimensional spin-$\frac{1}{2}$ gapped systems, to illustrate various technical 
aspects of the method and to discuss the physics of these topics. References to 
related specialised work dealing in more details with the physical 
aspects will also be given.

Low dimensional spin-$\frac{1}{2}$ systems with antiferromagnetic (AF) interactions display very innovative features, driven by strong quantum fluctuations. In particular, geometrical effects or competing magnetic interactions can give rise to the formation of a spin gap between the singlet ground state and the first excited triplet state. In this chapter, we focus on the numerical investigation of such systems by Exact Diagonalisation (ED) methods and some extensions of it including a simultaneous mean-field (MF) treatment of some perturbative couplings.\\
This Chapter is organised as follows: in Section 2 a description of the Lanczos algorithm is given with special emphasis on the practical use of space group symmetries. A very short review on the well-known planar frustrated Heisenberg model and some linear chain Heisenberg models is given in Section 3. In particular, we outline the role of the magnetic frustration in the formation of a disordered phase. We also introduce a MF treatment of interchain couplings. Section 4 is devoted to more recent studies of impurity doping and to the derivation of effective models describing interaction between dopants.

\section{Lanczos Algorithm}
\subsection{Algorithm}

The exact diagonalisation method is based on the Lanczos algorithm
\cite{Lanczos_ref} which we shall describe here.
This algorithm is particularly suited to handle sparse matrices and 
there is, in fact, a wide variety of lattice models belonging
to this class as we shall see later on~\cite{Haas}.  

Let us consider some lattice model corresponding to some Hamiltonian $H$ with 
its symmetry group ${\cal G}=\{ g\}$, namely $[H,g]=0$. 
Let us also assume, for the moment, that irreducible representations of the 
symmetry group can be constructed. They consist of 
complete subsets of states ${\cal A}_l=\{|\alpha\rangle\}$ which are globally 
invariant under the application of the Hamiltonian H
(we postpone to the next part of this section the explicit
construction of these states). Clearly, $H$ can be diagonalized 
in each of the subsets ${\cal A}_l$ independently. 
It can be written as a tridiagonal matrix in a new orthonormal basis set
$\{ |\Phi_m\rangle \}$ defined as~\cite{note_modified}
\begin{eqnarray} 
\nonumber
H|\Phi_1\rangle &=& e_1|\Phi_1\rangle +b_2|\Phi_2\rangle ,\\
                   &\vdots& \\  
\nonumber
H|\Phi_n\rangle &=& e_n|\Phi_n\rangle +b_{n+1}|\Phi_{n+1}\rangle
+b_{n}|\Phi_{n-1}\rangle \ ,
\label{Lanczos}                                 
\end{eqnarray}
where the various coefficients and new basis states 
can be calculated recursively. The proof is as follows:
let us suppose that the procedure has been applied until the step $n$, i.e.
an orthonormal set of states $|\Phi_{1}\rangle, ..., |\Phi_{n}\rangle$
has been constructed. 
Assuming the knowledge of $e_1,...,e_{n-1}$, $b_2,...,b_n$, 
$|\Phi_{n-1}\rangle$ and $|\Phi_{n}\rangle$, one can then 
determine 
\begin{equation}
e_{n}=\langle \Phi_{n}|H|\Phi_{n}\rangle  .
\end{equation}
Hence, the new state
defined by 
\begin{equation}
|\phi_{n+1}\rangle= H|\Phi_n\rangle - e_n|\Phi_n\rangle 
-b_{n}|\Phi_{n-1}\rangle  ,
\end{equation} 
is clearly orthogonal to $|\Phi_{n}\rangle$.
Moreover, $\langle\Phi_{n-1} |\phi_{n+1}\rangle
= \langle\Phi_{n-1}|H|\Phi_{n}\rangle-b_n$ which is also vanishing as can 
be seen by substituting the expression for $H|\Phi_{n-1}\rangle$. 
More generally, $|\phi_{n+1}\rangle$ is, in fact, orthogonal
to all the previous states $|\Phi_{p}\rangle$, $p\le n$ as can be shown
recursively. Indeed, let us assume that, for $p<n$, 
\begin{equation}
\forall i, \,\,\,\,\,  p\le i\le n \,\,\,\,\,\,
\langle\Phi_i|\phi_{n+1}\rangle =0, 
\end{equation}
then 
$\langle\Phi_{p-1}|\phi_{n+1}\rangle = \langle\Phi_{p-1}|H|\Phi_n\rangle $
where $\langle\Phi_{p-1}|\Phi_n\rangle 
=\langle\Phi_{p-1}|\Phi_{n-1}\rangle=0$ has been used.
Substituting the expression given by Eq. (\ref{Lanczos}) for $H|\Phi_{p-1}\rangle$
leads to the expected result 
\begin{equation}
\langle\Phi_{p-1}|\phi_{n+1}\rangle=0. 
\end{equation}
The (positive) number $b_{n+1}$ is simply defined as a normalisation
factor, 
\begin{equation}
b_{n+1}^2=\langle\phi_{n+1}|\phi_{n+1}\rangle ,
\end{equation}
i.e. $|\Phi_{n+1}\rangle=\frac{1}{b_{n+1}}|\phi_{n+1}\rangle$.

In principal, a zero vector will be generated after iterating the Hamiltonian 
a number of times corresponding to the size ${\cal N}_l$ of the 
Hilbert space. However, the number of iterations necessary to obtain the lowest
eigenvalues and eigenvectors is much smaller. Typically, 
for ${\cal N}_l\sim 10^6$, the ground state can be obtained with an accuracy better 
than $10^{-8}$, by truncating the procedure after only $N_{it}\sim 100$ 
iterations and by diagonalizing the resulting tri-diagonal matrix 
by using a standard library subroutine. 
However, for a given size ${\cal N}_l$ of the Hilbert space, 
the necessary number $N_{it}$ 
of iterations might vary by a factor of 2 or 3 depending on the model 
Hamiltonian. In practice, the convergence is faster for models for which
high energy configurations have been integrated out (e.g. t--J models 
in contrast to Hubbard models). 
Note however that, in some cases (models with strong finite range interaction),
the energy vs $N_{it}$ curve can exhibit steps before convergence to the 
true ground state is achieved.
Once space group symmetries have been implemented, the best choice for the 
initial state $|\Phi_1\rangle$ is a purely random vector. 
The ground state is also easily obtained as a function of the states $|\Phi_n\rangle$.
However, to express it in terms of the initial basis $\{|\alpha\rangle\}$
(as it is often useful) it becomes necessary to store temporarily 
the intermediate vectors $|\Phi_n\rangle$. This step is usually the most 
demanding in terms of disk space and/or mass storage.
However, note that, at runtime, only three Lanczos vectors of size 
${\cal N}_l$ need to be assigned in memory. 

\smallskip
\noindent {\underbar{\it Full diagonalisation:}}
In some special cases (spectrum statistics~\cite{chaos}, 
thermodynamics~\cite{thermodynamics}, etc ...) the complete spectrum
(or a least the lower part of it) is needed. This can also be performed
by the Lanczos algorithm. In this case, usage of a more sophisticated 
standard library package (e.g. EA15 of Harwell) is preferable.
Indeed, more involved tests become then necessary to eliminate the 
unphysical ``ghost'' levels appearing (always above the ground state energy)
after the diagonalisation of the
tridiagonal matrix. However, the input for the library subroutine 
consists only on the set of states $|\Phi_n\rangle$ which have to 
be calculated separately (see below). 


\subsection{Space group symmetries}

Usually, some efforts have to be carried out in order to take full 
advantage of the symmetries of the Hamiltonian. Let us consider a model defined
on a $D$-dimensional lattice describing interacting fermions
as e.g. the simple Heisenberg (spin) model,
\begin{equation}
H_J=\sum_{\bf x, y} J_{\bf x y}\, {\bf S}_{\bf x}\cdot {\bf S}_{\bf y},
\label{heisenberg}
\end{equation}
where interaction (in this case the exchange coupling) is not necessarily 
restricted to nearest neighbor (NN) sites but,
nevertheless, is supposed to exhibit translation and point group symmetry.
In other words, denoting the point group\cite{note1} by ${\cal G}_P=\{ g_P\}$,
we assume, e.g. in the case of Eq. (\ref{heisenberg}),  
\begin{eqnarray}
J_{\bf x y}&=&J({\bf x-y}) \nonumber \\
&\text{and}&  \\
\forall g_P&\in &{\cal G}_P, \,\,\,\, J(g_P({\bf r}))=J({\bf r}).
\nonumber 
\end{eqnarray}
Clearly, such properties are easily generalised to generic spin or 
fermionic Hamiltonians.  
In addition, we shall also assume, as in Eq. (\ref{heisenberg}),
spin rotational invariance 
(the total spin $S$ is a good quantum number) or, at least, invariance of the
Hamiltonian under a spin rotation around some quantisation axis.
Translation symmetry can be preserved on finite systems 
provided the geometry is that of a $D$-dimensional torus with periodic 
or twisted boundary conditions (BC).
On the torus geometry, the full space group reads,
\begin{equation}
{\cal G}={\cal G}_P\otimes {\cal T}  ,
\end{equation}
where we denote by ${\cal T}=\{ t_{p}\}$, $p=1,...,N$, the translation group. 
It is clear that $H_J$ is invariant under any $g\in{\cal G}$.

\begin{figure}
\centering
\psfrag{J1}{$J_1$}
\psfrag{J2}{$J_2$}
\psfrag{J+d}{$J+\delta$}
\psfrag{J-d}{$J-\delta$}
\psfrag{(a)}{(a)}
\psfrag{(b)}{(b)}
\psfrag{(c)}{(c)}
\includegraphics[height=3cm]{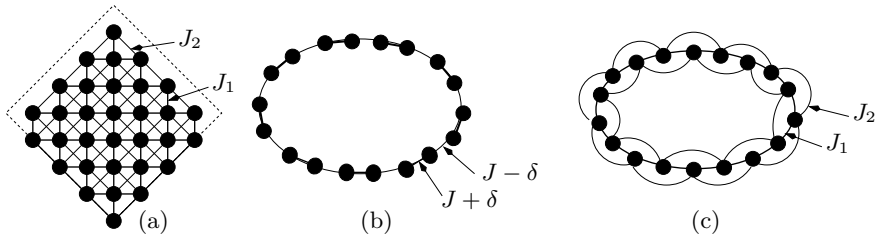}
\caption{Schematic representation of of different lattices. (a) the $\sqrt {32}\times \sqrt {32}$ square lattice with NN and second NN couplings $J_1$ and $J_2$. (b) the dimerized ring with dimerization $\delta$. (c) The frustrated ring with NN and second NN couplings $J_1$ and $J_2$.}
\label{lattices}       
\end{figure}
\subsection{Construction of the Hilbert space}

The motivation to take into account the Hamiltonian symmetries is two-fold.
First, the Hamiltonian can be block diagonalized, each block corresponding
to an irreducible representation of the symmetry group. Practically, the sizes
of the various blocks ${\cal N}_l$ are much smaller than the 
size of the full 
Hilbert space (see e.g. Table \ref{table1}), hence, minimising
the numerical effort to diagonalize the Hamiltonian matrix.
Secondly, each  irreducible representation of the symmetry group is 
characterised by quantum numbers (such as the momentum $\bf k$) which 
can be connected directly to physical properties. 

\begin{table}[htbp]
\centering
\caption{Symmetry groups and sizes of the reduced Hilbert spaces for
various spin-$\frac{1}{2}$ AF Heisenberg models for one of the most symmetric irreducible representation (typical reduced size is written as the total size in the $S^{z}=0$ sector divided by the number of symmetries). The 1D models are descibed in section \ref{pure.1d}. $T_N$ and $I_2$ stand 
for the translation group $\cal T$ (see text) and the spin inversion
symmetry $S_{\bf x}^Z \rightarrow -S_{\bf x}^Z$, respectively.}
\label{table1}
\begin{tabular}{|l|l|l|l|}
\hline{\smallskip}
Model & Lattice Size &  Symmetry group  & Typical reduced Hilbert space size\\
\hline
2D Isotropic &$6 \times 6$ & $T_{36}\otimes C_{4v}\otimes I_2$ & 9\, 075\, 135\, 300/576 \\
1D~$J_1 -J_2$ & $32 \times 1$ &  $T_{32}\otimes C_{2}\otimes I_2$  & 
601\, 080\, 390/128\\
1D~$J_1 -J_2 -\delta$ & $32 \times 1$ &  
$T_{16}\otimes I_2$ & 601\, 080\, 390/32\\
\hline
\end{tabular} 
\end{table}

We shall focus here on the Heisenberg model (\ref{heisenberg}) but t--J and Hubbard models can also be studied.
In order to minimise memory occupation, configurations can be stored in
binary form. One, first, assumes an arbitrary labelling of 
the lattice sites from 1 to L and denotes a configuration in real space by
\begin{equation}
|c\rangle =|s_1,...,s_i,...,s_L\rangle \,\, .
\label{config}
\end{equation}
For the Heisenberg model, the information $s_i$
on each lattice site ${\bf x}_i$ can be stored on a single bit, 
e.g. $\uparrow$ and $\downarrow$ 
correspond to 1 and 0 respectively. 
A spin configuration with up to 64 spins
can then be represented by a single 
64-bit word. As an example, a $N=4$ site configuration Heisenberg model
such as $| \uparrow,\downarrow,\uparrow,\downarrow\rangle$
is coded as $0_{64} ... 0_5 1_4 0_3 1_2 0_1 $ (where
the subscripts indicate the place of the bits) i.e. the integer ``10''.
An integer $N(|c\rangle)$ can then be uniquely associated to each 
configuration $|c\rangle$ and formally written as,
\begin{equation}
N(|c\rangle)=\sum_1^N 2^{i-1} \sigma_i \,\, ,
\label{code_Heis}
\end{equation} 
for spin-1/2 models ($\sigma_i=0$, or $1$).

At this point, it becomes useful to consider space group symmetry. 
Each irreducible representation can be characterised by a momentum 
\begin{equation} 
{\bf K}=\sum_\mu n_\mu {\bf K}_\mu \, ,
\end{equation}
where ${\bf K}_\mu$
are the reciprocal lattice vectors (e.g., in 2D,
${\bf K}_\mu=\frac{2\pi}{N}{\bf T}_\mu \wedge {\bf e}_z$)
and $n_\mu$ are 
integers. For each value of $\bf K$, one then considers ${\cal G}_{\bf K}^P$,
the so-called little group of $\bf K$ (${\cal G}_{\bf K}^P\subset{\cal G}_P$), 
containing group elements $g_P$ such that
\begin{equation} 
g_P({\bf K})={\bf K} \, .
\end{equation}
The relevant subgroup of ${\cal G}$ to be considered is then
\begin{equation} 
{\cal G}_{\bf K}={\cal G}_{\bf K}^P\otimes {\cal T} \, .
\end{equation}
For a given symmetry sector $l=({\bf K},\tau_{\bf K})$ ($\tau_{\bf K}$ 
is one of the irreducible representations of ${\cal G}_{\bf K}^P$)
a ``symmetric'' state $|\alpha\rangle$ can then be constructed from a 
single configuration $|c\rangle$ as a linear combination which reads, 
up to a normalisation factor,
\begin{equation}
|\alpha\rangle\equiv |\alpha\rangle \{|c\rangle\}=
\sum_{g_P\in{\cal G}_{\bf K}^P,t\in{\cal T}} 
e(\tau_{\bf K},g_P) \exp{(i{\bf K}\cdot{\bf T}_t)} \hspace{0.1cm}
(g_P \, t)(|c\rangle)  ,
\label{sym_state} 
\end{equation}
where $e(\tau_{\bf K},g_P)$ are the characters (tabulated) 
of the representation $\tau_{\bf K}$ and ${\bf T}_t$ are the translation 
vectors associated to the translations $t$. 
Since the procedure to construct the symmetric state is well defined, it is 
clear that one needs to keep only a single one of the related configurations
$g_P t(|c\rangle)$, this state being called ``representative'' of 
the symmetric state $|\alpha\rangle$ and denoted by 
$|r\rangle =R(|c\rangle)$. This naturally implies,
\begin{equation}
\forall g\in {\cal G}_{\bf K}, \,\,\, R(g(|c\rangle))=|r\rangle \, .
\end{equation}
In other words, one retains only the
configurations $|c\rangle$ which can not be related to each other
by any space symmetry
$g\in{\cal G}_{\bf K}$. The set of all the representatives (labelled from 
1 to ${\cal N}_l$) defines then unambiguously the Hilbert space 
${\cal A}_l=\{|\alpha\rangle\}$. Typically, the size of this 
symmetric subspace is reduced by a factor of $\text{card}({\cal G}_{\bf K})$ 
compared to the size of the original basis $|c\rangle$. Note that, in some
cases, there might exist some configurations $|c\rangle$ (to be eliminated) 
which do not give rise to any representative. This occurs when
there is a subset ${\cal G}^\prime \subset {\cal G}_{\bf K}$ such that
\begin{eqnarray} 
\forall g\in{\cal G}^\prime,& \,\,\, g(|c\rangle)=|c\rangle \nonumber \\
& \text{and}  \\
\sum_{g\in{\cal G}^\prime}e(\tau_{\bf K},g_P(g)) 
& \exp{(i{\bf K}\cdot{\bf T}(g))}\, =0 \, .  \nonumber
\end{eqnarray}

The choice of the representative among the set of equivalent states (i.e.
states related by some symmetry operations of ${\cal G}_{\bf K}$) is, in
principle, arbitrary. However, as we shall see in the following,
it is convenient to define it as the 
state $|c\rangle$ of a given class giving rise to the {\it smallest} 
integer $N(|c\rangle)$ i.e. 
\begin{equation}
N(|r\rangle )=\min_{g\in {\cal G}_{\bf K}} \{ N(g(|c\rangle)) \}
\end{equation}
For simplicity, we shall here extend our coding procedure to more general Hubbard-like models where one can construct the configurations 
by a tensorial product of the up and down spin parts 
\begin{equation}
|c\rangle=|c(\uparrow)\rangle\otimes |c(\downarrow)\rangle \, .
\end{equation}
$N(|c\rangle )$ contains now $2N$ bits and is of the form
\begin{equation}
N(|c\rangle )=N^\prime(|c(\uparrow)\rangle)\times 2^N 
+ N^\prime(|c(\downarrow)\rangle) \, ,
\label{coding}
\end{equation}
where $N^\prime(|c(\sigma)\rangle)$ corresponds to the binary coding 
($N^\prime(|c(\sigma)\rangle )<2^N$)) of the $\sigma$-spin part of the 
configuration, by using Eq.~(\ref{code_Heis}) where $1$ (resp. $0$) refers now to the presence (resp. absence) of a spin $\sigma$($=\uparrow$ or$\downarrow$) at a given site $i$. Although this coding is more costly in term of memory space ($2N$ bits per configuration instead of $N$), it is more general and applies both to Hubbard-like models (where $\uparrow$ and $\downarrow$ spins can leave on the same site) and to Heisenberg models. 
The minimisation 
of $N(g(|c\rangle))$ over $g\in {\cal G}_{\bf K}$ can be done in two steps. 
First, one generates all possible configurations for the up spins 
(this usually involves a limited number of states) and only representatives
$|r(\uparrow)\rangle$ are kept. At this stage, one needs to keep 
track of the subsets of symmetries 
${\cal E}_{\bf K}[r(\uparrow)\rangle]$ of ${\cal G}_{\bf K}$ leaving 
these representatives invariant. In a second step, one constructs the full 
set of configurations as a tensorial product of the form
$|r(\uparrow)\rangle\otimes |c(\downarrow)\rangle$.
The remaining symmetries of ${\cal E}_{\bf K}[|r(\uparrow)\rangle]$ 
are then applied to the 
spin $\downarrow$ part and one only retains the configurations 
$|c(\downarrow)\rangle$ such that,
\begin{equation}
\forall g^\prime \in {\cal E}_{\bf K}[r(\uparrow)\rangle] \,\,\,
N^\prime(|c(\downarrow)\rangle ) \le 
N^\prime [ g^\prime(|c(\downarrow)\rangle)]\, .
\end{equation}
The Hilbert space is then formally defined by all the symmetric 
states $|\alpha\rangle\{|r\rangle\}$ (see Eq.(\ref{sym_state})). 
However, one only needs to store the binary codes
$N(|r\rangle)$ as well as the normalisations of the related symmetric states. 
The normalisation of (\ref{sym_state}) requires some caution.
In some cases, there might exist more than a single (i.e the identity
operator $\mathbb{I}$) group element $g^\prime$ of 
${\cal E}_{\bf K}[|r(\uparrow)\rangle]$ which keeps $|r\rangle$ invariant.
Then, 
\begin{equation}
{\cal F}_{\bf K}[|r\rangle]
=\{ g^\prime\in {\cal E}_{\bf K}(|r(\uparrow)\rangle);
N^\prime[g^\prime(|c(\downarrow)\rangle)] =
N^\prime(|c(\downarrow)\rangle )\} 
\end{equation}
defines the subgroup of such symmetry operations.
The sum over the group elements in Eq. (\ref{sym_state}) should, in fact, be
restricted to the elements of the coset 
${\cal G}_{\bf K}/{\cal F}_{\bf K}[|r\rangle]$ and the appropriate 
normalisation factor is then given by
\begin{equation}
n(|r\rangle)=[ \frac{ \text{card}\{{\cal F}_{\bf K}[|r\rangle]\} }
{ \text{card}\{ {\cal G}_{\bf K}\} } ]^{1/2}\,\, .
\end{equation}
The two integers $N(|r\rangle)$ and $n(|r\rangle)$ 
corresponding to the binary code of a representative and to the 
normalisation of the related symmetric state, respectively, can be 
combined and stored in a single 64-bit computer word.

In the case of the generic Heisenberg model, the previous two-step 
procedure can, in most cases, also be done using the $N$ bits coding from Eq.(\ref{code_Heis}) but it is more involved. We only indicate here the spirit of the method.
More technical details for spin-1/2 models can be found
in Refs.~\cite{Schulz_J1J2} and \cite{Schulz_J1J2_2}.
The decomposition between up and down spins is replaced here by an
(appropriate) partition
of the lattice sites into two subsets $A$ and $B$. The computer word 
(integer) (\ref{code_Heis}) coding each
configuration contains then two parts, each part corresponding to one
subset of the lattice sites, $|c(A)\rangle$ and $|c(B)\rangle$. 
The partition is chosen in a way such that 
the group ${\cal G}_{\bf K}$ can be decomposed as 
\begin{equation}
{\cal G}_{\bf K}={\cal G}_{\bf K}^S + {\cal S}({\cal G}_{\bf K}^S) \, ,
\label{coset}
\end{equation}
where ${\cal S}$ is a group symmetry which fulfils ${\cal S}^2=\mathbb{I}$, 
the subgroup ${\cal G}_{\bf K}^S$ leaves 
the two subsets of lattice sites {\it globally} invariant and
${\cal S}({\cal G}_{\bf K}^S)$ is the coset of ${\cal S}$ relative to
${\cal G}_{\bf K}^S$. 
The decomposition (\ref{coset}) is not always unique. 
For 2D clusters such as
the $\sqrt{32}\times\sqrt{32}$ lattice of Fig.~\ref{lattices}, a convenient choice for $\cal S$ is a reflection symmetry.
Note that, in the case of 1D rings, (\ref{coset}) is only possible 
if the number of sites is of the form $N=4p+2$ or for very special
values of the momentum K (like $K=0$ or $K=\pi$).
Then, the previous procedure can be extended to this case by writing the 
configurations as
\begin{equation}
|c\rangle=|c(A)\rangle\otimes |c(B)\rangle
\label{conf_tJ}
\end{equation}
and by (i) applying the
subgroup ${\cal G}_{\bf K}^S$ on $|c(A)\rangle$ to generate its corresponding
representative $|r(A)\rangle$ and then (ii) applying all the symmetries 
of ${\cal E}_{\bf K}[|r(A)\rangle]$ on the part $|c(B)\rangle$.
The action of the remaining symmetry ${\cal S}$ is considered at last;
if the lattice sites are labelled in such a way that 
\begin{equation}
{\bf x}_{i+N/2}={\cal S}({\bf x}_{i}) \, , 
\end{equation}
for $i\le N/2$, the application of ${\cal S}$ 
can be implemented as a simple 
permutation of the two sub-words of $N(|c\rangle )$.

\subsection{Construction of the Hamiltonian matrix}

We turn now to the implementation of the basic operation 
$|\Phi_n\rangle\rightarrow H|\Phi_n\rangle$ 
appearing in Eq. (\ref{Lanczos}) which is always specific to the 
model Hamiltonian $H$ and constitutes the central part of the Lanczos code. 
Since the states $|\Phi_n\rangle$ are expressed in terms of the symmetric
states $|\alpha\rangle$ of (\ref{sym_state}) the Hamiltonian matrix
has to be computed in this basis. For this purpose, it is only necessary to
apply $H$ on the set of representatives 
$|r_\gamma\rangle$ (labelled from 1 to ${\cal N}_l$). In general, 
each configuration $|r\rangle$ leads to a small number $\beta_{max}$
(at most equal to the number of terms in H) of generated states,
\begin{equation}
H|r_\gamma\rangle \propto \sum_{\beta=1}^{\beta_{max}} 
(-1)^{\theta_{\gamma,\beta}} |c_{\gamma,\beta}\rangle \,\, ,
\label{matrix}
\end{equation}
where different signs $(-1)^{\theta_{\gamma,\beta}}$ might arise (in the case 
of fermion models) from fermionic commutation relations. 
The matrix is then very sparse. 
Note that the full Hamiltonian can always be split in a small 
number of separate terms so that 
the amplitude of the matrix elements in (\ref{matrix}) is just a 
constant and hence does not need to be stored.

At this point, it becomes 
necessary to determine the representatives (in binary form) 
of the various generated states on the right hand side of
(\ref{matrix}) by applying to them all the symmetries of ${\cal G}_{\bf K}$.
To achieve this, the choice of Eq. (\ref{coding}) for the 
binary coding of the configurations is very convenient.
It is, indeed, a simple way to take advantage of the
natural decomposition of the generated states,
\begin{equation}
 |c_{\gamma,\beta}\rangle=
|c_{\gamma,\beta}(\uparrow)\rangle\otimes |c_{\gamma,\beta}(\downarrow)\rangle \,\, .
\end{equation}
Although, we restrict ourselves here, for sake of simplicity, to the general coding of the
Hubbard-like models, the following procedure can be straightforwardly applied using the more restrictive Heisenberg form (\ref{conf_tJ}) for the configurations $|c_{\gamma,\beta}\rangle$.
The calculation of the representative
\begin{equation}
|r_{\gamma,\beta}\rangle_f=R\{ |c_{\gamma,\beta}(\uparrow)\rangle
\otimes |c_{\gamma,\beta}(\downarrow)\rangle\} 
\end{equation}
can be done in two steps. First, one applies
all the symmetries of ${\cal G}_{\bf K}$ to 
$|c_{\gamma,\beta}(\uparrow)\rangle$.
Since this procedure has to be repeated a large number of times,
the function
\begin{equation}
R:\ |c(\uparrow)\rangle\, \longmapsto\,
|r(\uparrow)\rangle \, .
\end{equation}
can be, in fact, tabulated, prior to the actual calculation of the
matrix elements. This is made possible since the number of possible states
$|c(\uparrow)\rangle$ remains, in general quite modest.
This procedure enormously speeds up the calculation of the representatives
and justifies the choice of Eq.~(\ref{coding}). 
Note that one also needs, in this preliminary calculation, to 
store, for each configuration $|c(\uparrow)\rangle$, 
the corresponding ensembles,
\begin{equation}
{\cal R}_{\bf K}[(|c(\uparrow)\rangle]
=\{g\in {\cal G}_{\bf K} ; g(|c(\uparrow)\rangle)= |r(\uparrow)\rangle\}\, .
\end{equation}
This also requires limited storage since, in most cases,
${\cal R}_{\bf K}[|c(\uparrow)\rangle]$ contains just a single element.
In a second step, it is sufficient to apply only the remaining
symmetries of ${\cal R}_{\bf K}[|c_{\gamma,\beta}(\uparrow)\rangle]$ to the 
$|c_{\gamma,\beta}(\downarrow)\rangle$ part. 
A standard hashing table~\cite{hashing} is then used to find 
the positions of the representatives in the list. 
The connectivity matrix connecting the labels of the initial set of 
representatives to the labels of the new set of generated states is then 
stored on disk.

Note that the phases related to commutations of fermion operators and/or 
to the characters of the symmetry operations involved in the transformation 
of the generated states to their representatives have also to be kept.
These phases have the general form,
\begin{equation}
\lambda_{\gamma,\beta}=(-1)^{\theta_{\gamma,\beta}} 
e(\tau_{\bf K},g_P(g^*_{\gamma,\beta})) \exp{(i{\bf K}
\cdot{\bf T}(g^*_{\gamma,\beta}))}\,\, ,
\label{phase}
\end{equation}
where $g^*_{\gamma,\beta}$ is a group element of 
${\cal R}_{\bf K}[|c_{\gamma,\beta}(\uparrow)\rangle]$ such that
\begin{equation}
|r_{\gamma,\beta}\rangle_f= g^*_{\gamma,\beta}
(|c_{\gamma,\beta}\rangle) \,\, ,
\label{optimal_g}
\end{equation}
which depends on $\gamma$ and $\beta$. 
It is easy to show that, if there exists more than a single group element
which fulfils (\ref{optimal_g}) then, all of them lead to the same 
phase factor (\ref{phase}). 

Since the number of possible 
different phases given by (\ref{phase}) is quite small, it is possible 
to store the $\lambda_{\gamma,\beta}$ (in some convenient integer form)
together with $N\{|r_{\gamma,\beta}\rangle_f\}$ 
on a small number of computer bits. For an Hilbert space of $10^8$ 
(hundred millions) representatives with typically an average of
$\sim 50$ images per state,
the occupation of the disk is of the order of 5000 Mw i.e. 40 Gb.
This can even be reduced by a factor of two 
by using each computer word to store the
informations corresponding to two images instead of a single one. 
Once it has been generated, the Hamiltonian matrix 
is cut into several pieces (typically of the order of 10 to 100 Mw) and 
the various parts are successively read from the disk in order 
to calculate $H|\Phi_n\rangle$. The best performances are obtained when the
calculation using the n$^{th}$ part of the matrix and the access to the disk 
to read the (n+1)$^{th}$ part are simultaneous. 
Note that the Lanczos algorithm as it has been described 
above is well adapted to be implemented on a vector 
supercomputer (e.g. on a NEC-$5\times 5$). 

\section{Examples of translationally invariant spin gapped systems}
Here, we first restrict ourselves to systems where the symmetry analysis decribed above can be used. Note however that explicit symmetry breaking may be present but, in general, the remaining symmetry group contains a large number of symmetries which can be exploited. Note also that since spontaneous symmetry breaking can occur only in the thermodynamic limit, it does not prevent for finite systems the previous symmetry analysis.
\subsection{Application to the $2$D $J_1-J_2$ model}

The study of quantum phase transitions in 2D is of great interest. 
One of the standard example is the so-called frustrated Heisenberg model (see Fig.\ref{lattices}(a)).
It is defined by (\ref{heisenberg}) when only NN and second NN exchange couplings are retained,
\begin{eqnarray}
J_{\bf x y}&=&J_1\ \text{if}\ |{\bf x-y}|=1, \nonumber \\
J_{\bf x y}&=&J_2\ \text{if}\ |{\bf x-y}|=\sqrt{2},    \\
J_{\bf x y}&=&0\ \text{otherwise}\, . \nonumber
\end{eqnarray}
Various analytical approaches have been applied to this problem
such as spin-wave calculations~\cite{spin_wave_J1J2}, 
large-N SU(N) theories~\cite{large_N}, series expansions~\cite{series_J1J2}, 
and Schwinger boson mean field approaches~\cite{Schwinger_J1J2}.
However, most of these methods are somewhat biased since they assume 
the existence of some particular ground states. Pioneering unbiased ED 
studies~\cite{Dagotto_J1J2,other_J1J2,Didier_J1J2} 
have strongly suggested the existence of a disordered magnetic phase for
intermediate couplings $J_2/J1$. Here, we briefly discuss some more recent
subsequent work~\cite{Schulz_Ziman,Schulz_J1J2,Schulz_J1J2_2}
which attempted to obtain more accurate results by 
a finite size scaling analysis. Similar studies have also been performed
for other $S=1/2$ 2D spin models like the 
triangular lattice~\cite{Bernu_triangulaire},
the Kagome lattice~\cite{Bernu_Kagome}, the 1/5-depleted square
lattice~\cite{CaVO} or the (2D) pyrochlore lattice~\cite{pyrochlore}.

\begin{figure}
\centering
\includegraphics[height=7cm]{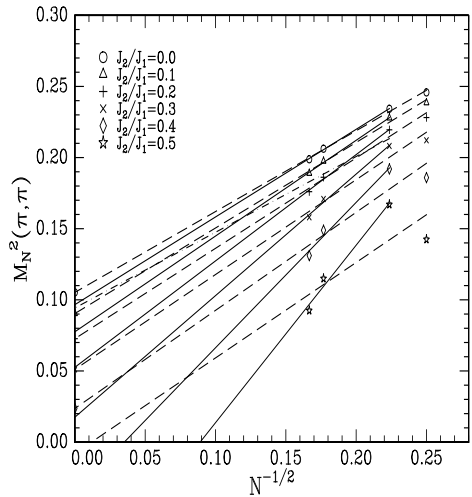}
\caption{Finite size results for $M_N^2(\pi,\pi)$ for
different
values of $J_2$. The dashed lines are least squares fits to the data, using all available clusters. 
The full lines are fits using only $N=20,32,36$ (Reprinted from 
Ref.~\protect\cite{Schulz_J1J2_2}).
}
\label{J1J2_1}
\end{figure} 

A magnetic ordered phase with a spin modulation $\bf Q$ can be characterised by
an order parameter $M_N({\bf Q})$ defined by
\begin{equation}
M_N^2({\bf Q})=\frac{1}{N(N+2)} \langle\Psi_0|
(\sum_{i=1}^N \exp{(i\, {\bf x}_i\cdot {\bf Q})}\, {\bf S}_i )^2 
|\Psi_0\rangle\ ,
\label{magnetisation}
\end{equation}
where $|\Psi_0\rangle$ is the ground state. It is important to notice that, in any finite
system, the order parameter itself has a zero expectation value in the ground state
due to spin SU(2) symmetry. In other words, the macroscopic magnetisation 
can slowly fluctuate so that in average it vanishes. It is therefore essential
to consider, as in (\ref{magnetisation}), the square of the order parameter
which can be interpreted as a generalised susceptibility. 

For weak frustration $J_2/J_1$, N\'eel order with ${\bf Q}=(\pi,\pi)$
is expected, while for large ratio $J_2/J_1$ a collinear phase with
${\bf Q}=(\pi,0)$ or ${\bf Q}=(0,\pi)$ consisting of successive 
alternating rows of parallel spins is a serious candidate.
Indeed, in such a collinear phase, each sub-lattice has N\'eel order so it 
is clear that it is stabilised by $J_2$.
Note that the normalisation factor of the staggered magnetisations (\ref{magnetisation}) is chosen so that the order parameter is independent of the size in a
perfect classical N\'eel or collinear state.
In such ordered phases where the continuous spin symmetry is spontaneously
broken, field theory arguments~\cite{sigma_model} suggest a scaling of the
form,

\begin{equation}
M_N({\bf Q})\simeq m_0({\bf Q}) +\frac{C({\bf Q})}{N^{1/2}}\  .
\label{scaling2}
\end{equation}

\begin{figure}
\centering
\includegraphics[height=6cm]{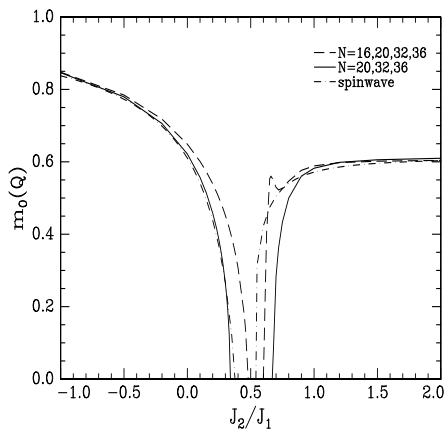}
\caption{ Comparison of the finite size fits for the anti-ferromagnetic and
collinear order parameters (left and right curves, respectively) with
linear spin wave theory (Reprinted from
Ref.~\protect\cite{Schulz_J1J2_2}).}
\label{J1J2_2}
\end{figure}

Square clusters with $N=4p$ sites (so that $(\pi,0)$ belongs to the
reciprocal lattice) are considered (see Sec. II.B), i.e. 
N=16, 20, 32 and 36. 
As seen in Fig.~\ref{J1J2_1} corresponding to ${\bf Q=(\pi,\pi)}$ 
the scaling law (\ref{scaling2}) is very well satisfied. 
Note that the $4\times 4$ cluster shows systematic deviations.
Similar results are also obtained for the collinear order parameter
at larger $J_2/J_1$ ratios.

The extrapolated results are shown in Fig.~(\ref{J1J2_2}). 
The most interesting feature is the existence of a narrow range of
$J_2/J_1$ around $0.5$ where none of the ordered states is stable. 
Various candidates for this disordered phase have been proposed
such as the dimer phase~\cite{large_N} and investigated 
numerically~\cite{Dagotto_J1J2,Didier_J1J2,Schulz_Ziman}.
More details on this topic can be found e.g. in the review 
article~\cite{Schulz_J1J2}.

\subsection{Application to spin-Peierls chains}

\paragraph{a) Purely 1D models}
\label{pure.1d}
We now move to systems with anisotropic couplings in space (quasi 1D materials). Let us first consider purely 1D models in order to describe the physical origin of the spin-Peierls (SP) transition. One of the simplest 1D model which diplays such a behavior is the frustrated spin-$\frac{1}{2}$ ring (see Fig.\ref{lattices}(c)), also called $J_1 -J_2$ or zig-zag chain, described by the Hamiltonian
\begin{equation}
\label{1d.frust}
H_{\rm{frust}}=\sum_{i=1}^{L}(J_1\vec{S}_i.\vec{S}_{i+1}+J_2 \vec{S}_i.\vec{S}_{i+2}).
\end{equation}
Its symmetry properties are given in Table \ref{table1} for $L=32$ sites.

\begin{figure}
\centering
\includegraphics[height=6cm]{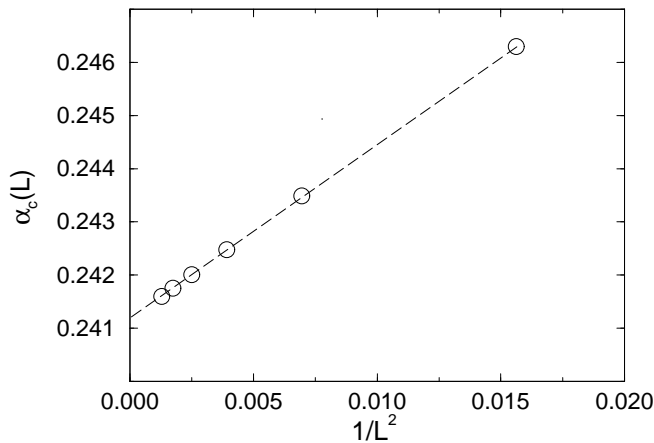}
\caption{Critical value $\alpha_c$ of the frustration vs the inverse square of the system size obtained using the method developped in \cite{emery88}. (Reprinted from Ref.\cite{theseaugier}).}
\label{alphac.L}
\end{figure}

The low energy properties of such a model are very interesting because it is gapless as long as $\alpha=J_2/J_1$ remains smaller than a critical value $\alpha_{c}\simeq 0.2412$~\cite{Eggert96}. For $\alpha > \alpha_c$, a gap $\Delta_{S}(\alpha)\propto e^{-(\alpha-\alpha_c)^{-1}}$ develops and a spontaneous dimerization appears, characteristic of the SP transition. At $\alpha=0.5$, the so-called Majumdar-Ghosh (MG) point~\cite{MG}, the 2-fold degenerate ground state is known exactly and consists in the product of spin singlets located either on odd or on even bonds. Beyond the MG point, the short-range correlations become incommensurate. The triplet ($S=1$) spectrum of the SP phase is a two-particle (so-called kink or soliton) continuum as evidenced by the scaling of the soliton-antisoliton binding energy to zero~\cite{augier. phonons.98}.\\

Adding an explicit dimerization , i.e. a rigid modulation $\delta$ of the NN coupling (see Fig.\ref{lattices}), drives immediately the ground state into a SP phase for any $\delta \ne 0$ even if $J_2=0$. In term of symmetries, $C_2$ and the translations of an odd number of lattice spacings are lost (see Fig.\ref{lattices}(b) and Table \ref{table1})). If $J_2 \ne 0$, the dimerized {\it and} frustrated model
\begin{equation}
\label{1d.dim.frust}
H_{\rm{dim}}=\sum_{i=1}^{L}J_1\left[(1+(-1)^{i}\delta)\vec{S}_i.\vec{S}_{i+1}+\alpha \vec{S}_i.\vec{S}_{i+2}\right]
\end{equation}
displays an enhancement of the dimerized gapped phase, indeed $\Delta_S(\alpha,\delta)-\Delta_S(\alpha,0)\propto \delta^{2/3}$~\cite{delta2.3}. 

\begin{figure}
\centering
\psfrag{0}{$0$}
\psfrag{Gapless}{{\bf gapless}}
\psfrag{alphac}{{\small{$\alpha_c\simeq 0.2412$}}}
\psfrag{1}{$1$}
\psfrag{d}{$\delta$}
\psfrag{Gapped}{{\bf Gapped}}
\psfrag{SP}{{\bf SP}}
\psfrag{commensurate}{{\bf Commensurate}}
\psfrag{incommensurate}{{\bf Incommensurate}}
\psfrag{0.5}{$0.5$}
\psfrag{MG}{MG}
\psfrag{alpha}{$\alpha$}
\includegraphics[height=4cm]{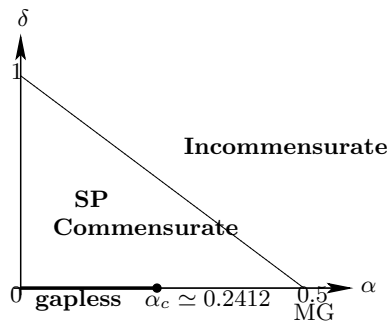}
\caption{Phase diagram of the frustrated dimerized Heisenbeg AF spin-$\frac{1}{2}$ chain in the $\alpha-\delta$ plane. The dotted line is the Shastry-Sutherland line ($2\alpha+ \delta=1$)~\cite{s.s.line}. On its left side, the phase is SP gapped and commensurate whereas on the right side, the correlations are incommensurate.}
\label{phasdiag.1d}
\end{figure}
\begin{figure}
\centering
\psfrag{L=32,k=0}{$L=32,k=0$}
\psfrag{k}{$k$}
\psfrag{E/J}{$\frac{E}{J}$}
\includegraphics[height=6cm]{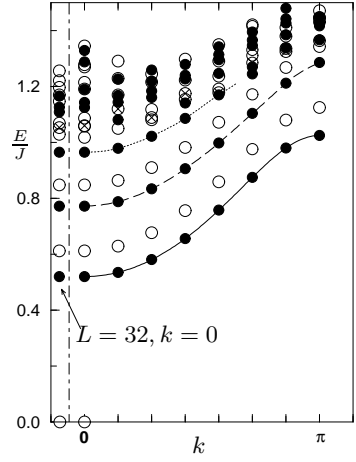}
\caption{Lowest lying triplet($\bullet$), singlet ($\circ$) and quintuplet ($\otimes$) excitations vs. the wave vector $k$ for the dimerized frustrated chain Eq.(\ref{1d.dim.frust}) with $J_2=0.5,~\delta=0.05,~L=28$. Results for $L=32,~k=0$ are shown to the left. ED results reprinted from~\cite{Sorensen98}.}
\label{fig2.augier}
\end{figure}

These properties are summarized in the phase diagram shown in Fig.\ref{phasdiag.1d}. Note that the static modulation $\delta$ leads to soliton-antisoliton boundstaes as shown in Fig.\ref{fig2.augier}.

\begin{figure}
\centering
\psfrag{J}{$J_{\bot}$}
\psfrag{AF}{AF}
\psfrag{SP}{SP}
\psfrag{d}{{\tiny{$L=12$}}}
\psfrag{b}{{\tiny{$L=16$}}}
\psfrag{m}{{{$m$}}}
\psfrag{pJ}{{\tiny{$J_{\bot}$}}}
\psfrag{MG}{{\tiny{MG}}}
\includegraphics[height=6cm]{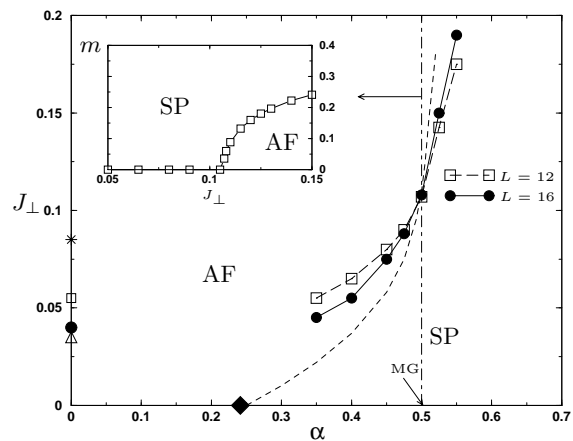}
\caption{Phase diagram of the coupled frustrated chains as a function of the frustration $\alpha$ and the inter-chain magnetic coupling $J_{\bot}$. The points, calculated for 2 different chain sizes (12 and 16 sites), separate a dimerized phase (SP) from a N\'eel ordered phase (AF). The closed diamond shows the order-disorder critical point at $\alpha _c\simeq 0.2412$. The dashed line represents the expected behavior in the thermodynamic limit. In the inset, the staggered magnetization $m(J_{\bot})$ has been calculated for a $L=12$ sites chain along the MG line (dot-dashed line).
Along the $\alpha=0$ line, different symbols show the critical $J_{\bot}$ for $L=8, 12, 16, 18$ from top to bottom and we have checked its scaling to $0$ according to a $1/L$ law.}
\label{fig:PhDg1}
\end{figure}

\paragraph{b) Chain mean field theory for coupled spin chains}
Physically, the previous models are often inadequate to describe the properties of several compounds like CuGeO$_3$\cite{Hase93} or LiV$_2$O$_5$\cite{LiV2O5} which are excellent realizations of weakly interacting frustrated spin-$\frac{1}{2}$ chains.
Let us first consider a set of frustrated spin chains which are coupled by a weak AF exchange $J_{\perp}$. This 2D model is governed by the following Hamiltonian
\begin{eqnarray}
\label{2d}
H_{2D}(\alpha,J_{\perp})= \sum_{i=1}^{L}\sum_{a=1}^{M}[\vec{S}_{i,a}.\vec{S}_{i+1,a}+\alpha \vec{S}_{i,a}.\vec{S}_{i+2,a} \nonumber \\
+J_{\perp}{\vec S}_{i,a}.{\vec S}_{i,a+1}].
\end{eqnarray}
where $i$ is the lattice index along the chains of lenght $L$ and $a$ labels the $M$ chains ($L$ and $M$ are chosen to be even and periodic boundary conditions are assumed in both directions).
Obviously it should be possible to study exactly this spin model on the square lattice but, as we have seen above, it is hard to perform ED with system larger than $36$ spins. Here, we take advantage of the fact that $J_{\perp}<<1$ to perform a MF treatment of the transverse coupling.  
Following Schulz~\cite{Schulz96}, the chain mean-field (CMF) version of (\ref{2d}) is given by 

\begin{eqnarray}
\label{2d.mf}
H_{2D}^{\rm MF}(\alpha,J_{\perp})= \sum_{i=1}^{L}\sum_{a=1}^{M}[\vec{S}_{i,a}.\vec{S}_{i+1,a}+\alpha \vec{S}_{i,a}.\vec{S}_{i+2,a} \nonumber \\
+h_{i,a}S^{z}_{i,a}-J_{\perp}{\langle S}_{i,a}^{z}\rangle{\langle S}_{i,a+1}^{z}\rangle],
\end{eqnarray}
with
\begin{equation}
\label{Hi}
h_{i,a}=J_{\bot}(\langle S_{i,a+1}^z \rangle+\langle S_{i,a-1}^z \rangle), 
\end{equation}
the local magnetic field to be computed self-consistently. In the
absence of dopant (see Section 4), we expect an homogeneous AF phase characterized
by a self-consistent staggered magnetization $\langle
S^{z}_{i,a}\rangle=(-1)^{i+a} m$. 
Therefore the coupled chains problem is reduced to a single chain in a staggered magnetic field $h_i=\pm 2(-1)^i J_\perp m$. 
\begin{equation}
\label{singlechain}
H_{\rm single} (\alpha,J_{\perp})=\sum_{i=1}^{L}[{\vec S_{i}}.{\vec S_{i+1}}+\alpha {\vec S_{i}}.{\vec S_{i+2}}+2mJ_{\perp}(-1)^i S_{i}^{z}]+{\rm constant},
\end{equation}
and the symmetry group of such a model is $T_{L/2}$.
In the absence of frustration ($\alpha=0$), it was
shown that $m\sim \sqrt{J_{\bot}}$~\cite{Schulz96}. By solving the
self-consistency condition using ED of finite chains the
transition line $J_{\bot}=J_{\bot}^c(\alpha)$ (see
Fig.~\ref{fig:PhDg1}) separating the  dimerised SP phase ($m=0$)
and the AF ordered phase (for which $m\neq 0$) has been obtained
in agreement with field theoretic approaches~\cite{Fukuyama96}.
Finite size effects are small in the gapped regime and especially
at the MG point. Note also that numerical data suggest that the AF order
sets up at arbitrary small coupling when $\alpha < \alpha_c$ with
a clear finite size scaling $J_{\bot}^c(L)\propto 1/L$ at small
$\alpha$.

\paragraph{c) Convergence issues of the numerical CMF}

\begin{figure}
\centering
\psfrag{delta}{$\delta J_{\perp}$}
\psfrag{xi}{$\xi_{\tau}$}
\psfrag{p}{$t$}
\psfrag{m}{$\ln|V(t)|$}
\psfrag{SP}{AF}
\psfrag{AF}{SP}
\includegraphics[height=6cm]{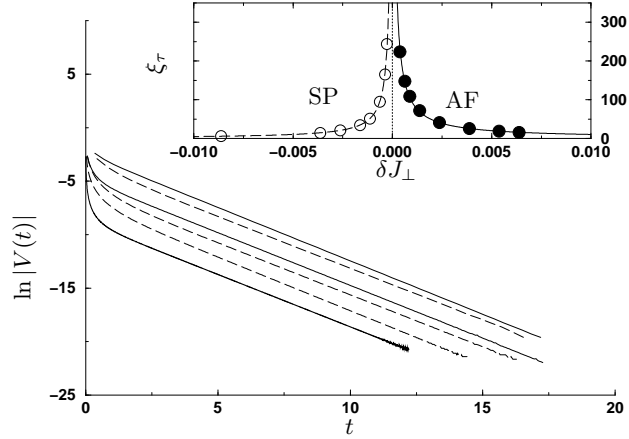}
\caption{Universal behavior of the convergence speed of the MF iterative procedure plotted versus the renormalized iteration index $t=\frac{p}{\xi_{\tau}}$. Results are shown for a system of spins interacting via Eq.(\ref{singlechain}) with $L=12$ and $\alpha=0.5$. From top to bottom $J_{\perp}=0.075, 0.13, 0.1, 0.11, 0.108, 0.106$. An initial value $m(0)=1/2$ is used for all simulations. Convergences to $m^*=0$ if the phase is SP (solid lines) or $m^* \ne 0$ (long-dashed lines) if the pase is AF are obtained. The inset shows the behavior of the typical convergence time scale $\xi_{\tau}$ as a function of the distance to the critical point $\delta J_{\perp}=J_{\perp}-J_{\perp}^{c}$. The curves are power law fits (see text).} 
\label{time.cvg}
\end{figure}

The numerical procedure consists of successive Lanczos-diagonalizations of a frustrated ($\alpha)$ spin-$1/2$ chain. At each step the AF order parameter $m$ is calculated and reinjected at the following step as the ``new field''.
Starting the numerical procedure with an arbitrary value of $m(0) \ne 0$, the chain of size $L$ is first diagonalized, $m(1)$ is extracted and then used for the next iteration. Eventually the procedure converges to the fixed point $m^*$.
A very interesting feature is that the convergence to $m^*$ as function of the number of MF iterations $p$ (see Fig.\ref{time.cvg}) is exponential.
\begin{equation}
\label{exp.cvg}
m(p)-m^* \propto  \exp(-p/\xi_{\tau})~~~{\rm for~} p>>\xi_{\tau},
\end{equation}
with $\xi_{\tau}(J_{\perp})$ a typical convergence time scale.\\
In order to study convergence at large p we have considered here small systems ($12$ sites) at the MG point where the finite size effects are very small.
We have checked this convergence issue by studing the speed $V(p)=|m(p+1)-m(p)|$ which is exponentially vanishing
\begin{equation}
\label{speed.cvg}
V(p)\propto \exp(-t)~~~{\rm for~} t=\frac{p}{\xi_{\tau}}>>1,
\end{equation}
as we can see in Fig.~\ref{time.cvg}. It is very important to note that the convergence of the MF procedure is universal in the sense that the choice of the starting value $m(0)$ is not crucial.\\
The time scale $\xi_{\tau}$ is $J_{\perp}$-dependant and diverges as a power law $\xi_{\tau}\sim |\delta J|^{-\mu}$ when approaching the critical line where $|\delta J|=|J_{\perp}-J_{\perp}^{c}|$. Our datas suggest $\mu\simeq 1.06~\rm{if}~J_{\perp}^{c}>J_{\perp}$ and $\mu\simeq 0.95~\rm{if}~J_{\perp}>J_{\perp}^{c}$ for $L=12$ (see inset of Fig.~\ref{time.cvg}).


\section{Lanczos algorithm for non uniform systems~:~Application to doped SP chains}
\label{sec:4}

Doping a SP system with non-magnetic impurities leads to very surprising new features. For example in Cu$_{1-x}$M$_x$GeO$_3$ (M$=$Zn or Mg), the discovery of coexistence between dimerization and AF long range order at small impurity concentration has motivated extented experimental~\cite{Zn_CuGeO} and theoretical~\cite{Sorensen98,Nakamura99,Martins96,Normand2002,Hansen99,Dobry98} investigations. In the following we report numerical studies of models for doped coupled spin chains. For sake of completness we also include a four-spin coupling which originates from cyclic exchange~\cite{cyclic,Laeuchli2002}.  

\subsection{Doped coupled frustrated spin-$\frac{1}{2}$ chain with four-spin exchange}

As for the transverse coupling $J_{\perp}$ in Eqs.(\ref{2d.mf},\ref{Hi}), we also apply the MF treatment to the added $4$-spin coupling $J_4(\vec S_{i,a}\cdot\vec S_{i+1,a})(\vec
S_{i+1,a+1}\cdot\vec S_{i,a+1})$. This leads to a self-consistent modulation of the NN couplings

\begin{eqnarray}
\label{hamilQ1D.MF} H_{\rm eff}(\alpha,J_{\bot},J_4)
=\sum_{i,a}[(1+\delta J_{i,a})\vec S_{i,a}\cdot \vec S_{i+1,a}
\nonumber \\ + \alpha\vec S_{i,a}\cdot \vec
S_{i+2,a}+h_{i,a}S_{i,a}^z] + {\rm constant}\, ,
\end{eqnarray}

with  $h_{i,a}$ given by Eq.(\ref{Hi}) and

\begin{equation}
\label{delti} \delta J_{i,a}=J_4\lbrace\langle \vec
S_{i,a+1}\cdot\vec S_{i+1,a+1} \rangle+\langle \vec
S_{i,a-1}\cdot\vec S_{i+1,a-1} \rangle\rbrace.
\end{equation}
Such a modulation produced by the $J_4$ term stabilizes the SP phase and raises the transition line in Fig.\ref{fig:PhDg1} (for more details, see \cite{ourprl03}). Another interesting feature of this model is its direct link with the magneto-elastic model considered in \cite{Hansen99} where the elastic coupling $K$ plays a role very similar to that of $1/J_4$. In the following the parameters $\alpha, J_{\perp}$ and $J_4$ are set in order to constrain the system to be in a SP state in the absence of dopants.\\ 

\begin{figure}
\centering
\psfrag{J1}{{\small{$J_1$}}} \psfrag{alpha J1}{{\small{$\alpha J_1$}}}
\psfrag{Jp}{{\small{$J_{\bot}$}}} \psfrag{J4}{{\small{$J_4$}}}
\includegraphics[height=4cm]{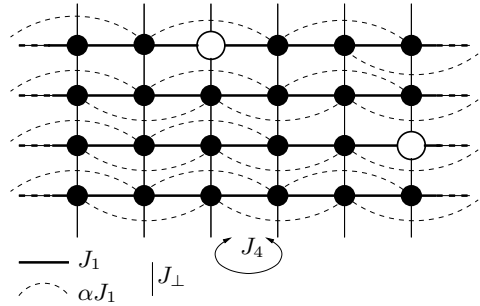}
\caption{Schematic picture
of the coupled chains model with nearest neighbor, next-nearest
neighbor, inter-chain and $4$-spin couplings $J_1$, $J_2=\alpha
J_1$, $J_{\bot}$, and $J_4$. Full (resp. open) circles stand for
spin-$\frac{1}{2}$ sites (resp. non-magnetic dopants).}
\label{fig:lattice.cpled}
\end{figure}

A dopant is described as an inert site i.e. all couplings to this site will be set to zero. Contrary to what we have seen previously, the use of the translation invariance is now forbidden by the presence of a single defects or by randomly located defects (see Fig.\ref{fig:lattice.cpled}). The maximal size accessible with a Lanczos procedure is then reduced because of this lack of symmetry and also because of the repeated iterative MF procedure. Indeed, the problem can not be reduced to a single chain model and hence the $M$ {\it non-equivalent} chains have to be diagonalized independently. Following the method used in Ref.\cite{Dobry98}, the MF equations are solved self-consistently on finite $L\times M$ clusters. Therefore, in the doped case, the time scale of the MF convergence $\xi_\tau$ for the single chain problem (\ref{singlechain}) is typically multiplied by $M$.
\subsection{Confinement}

 Replacing
a single spin-$\frac{1}{2}$ in a {\it spontaneously} dimerized (isolated)
spin chain by a non magnetic dopant (described as an inert site)
liberates a free spin $\frac{1}{2}$, named a soliton, which does not
bind to the dopant~\cite{Sorensen98}. The soliton can be depicted
as a single unpaired spin (domain) separating two dimer
configurations~\cite{Sorensen98}. The physical picture is
completely different when a {\it static} bond dimerisation exists
and produces an attractive potential between the soliton and the
dopant~\cite{Sorensen98,Nakamura99} and consequently leads, under
doping, to the formation of local magnetic
moments~\cite{Sorensen98,Normand2002} as well as a rapid
suppression of the spin gap~\cite{Martins96}. However, a coupling
to a purely one-dimensional (1D) adiabatic lattice~\cite{Hansen99}
does not produce confinement in contrast to more realistic models
including an elastic inter-chain coupling (to mimic 2D or 3D
lattices)~\cite{Hansen99,Dobry98}.\\
Here, we re-examine the confinement problem in the context of the previous model including interchain magnetic coupling.
\paragraph{a) Different kinds of dimer orders}

\begin{figure}
\centering
\psfrag{J4}{$J_4$}
\psfrag{Delta}{$\Delta_{\rm Stg-Clm}$}
\includegraphics[height=5cm]{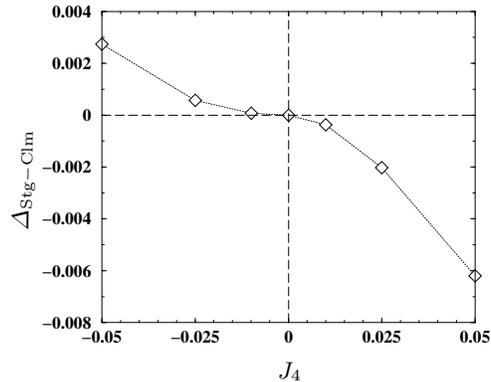} 
\caption{Energy difference $\Delta_{\rm Stg-Clm}$ between ground states with staggered and collumnar dimer orders plotted versus $J_4$ for model (\ref{hamilQ1D.MF}) (without dopant) at $\alpha=0.5,~J_{\perp}=0.1$ and $L=12$.}
\label{fig:gap.confin}
\end{figure}

\begin{figure}
\centering
\psfrag{a}{(a)~~$J_4 =0$}
\psfrag{b}{(b)~~$J_4 <0$}
\psfrag{c}{(c)~~$J_4 >0$}

\includegraphics[height=3cm]{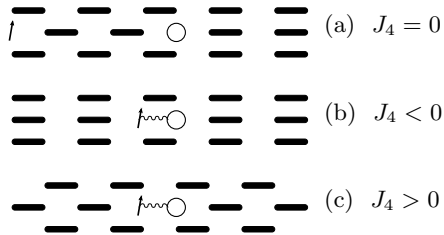} 
\caption{Schematic picture of the soliton confinement mechanism induced by the coupling $J_4$ of the model (\ref{hamilQ1D.MF}). The non magnetic dopant is represented by an open circle and the large black bonds stand for stronger dimer bonds. The black arrow represents the soliton, released by the impurity, which is deconfined if $J_4=0$ (a) whereas it is linked to the dopant if $J_4\ne 0$. We can see that this binding is imposed by the bulk dimerization which is columnar if $J_4 <0$ (b) or staggered if $J_4>0$ (c)}
\label{fig:confin.pict}
\end{figure}

\begin{figure}
\centering
\psfrag{M1}{${\mathcal{M}}_{i,a}^{\rm stag}$} \psfrag{x}{$i$}
\psfrag{a}{{\small{$J_4=0$}}} \psfrag{b}{{\small{$J_4=0.01$}}}
\psfrag{c}{{\small{$J_4=0.08$}}}
\includegraphics[height=6cm]{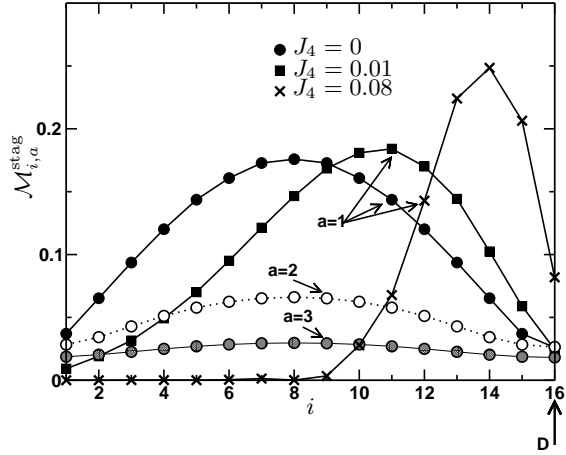} 
\caption{Local
magnetization ${\mathcal{M}}^{\rm stag}_{i,a}$ for $L\times
M$=$16\times 8$ coupled chains with one dopant D (shown by arrow)
located at $a=1$, $i=16$ in the dimerised phase ($\alpha=0.5$,
$J_{\bot}=0.1$). Circles correspond to $J_4=0$ (shown up to the
third neighbor chain of the doped one) and squares (crosses) to
$J_4=0.01$ ($J_4=0.08$). The coupling $J_2$ across the dopant has been
set to $0$ for convenience (Reprinted from Ref.\cite{ourprl03}).} 
\label{fig:ConfinJ2p}
\end{figure}

Let us return to model (\ref{hamilQ1D.MF}). For $J_4=0$, the MF treatment of the transverse magnetic coupling $J_{\perp}$ does not break the degeneracy of the ground state~:~each chain displays a $2$-fold degenerate ground state (dimers can stand either on even or odd bonds) independently from the other ones. The situation changes radically when $J_4\ne 0$ because the degeneracy is reduced to $2$. Indeed, each chain displays the same dimerized pattern if $J_4<0$ (columnar dimer order) whereas the dimer order is staggered  in the transverse direction if $J_4 >0$, as we can see in Fig.\ref{fig:gap.confin}.
Consequently, the soliton remains deconfined~\cite{Byrnes99} when $J_{4}=0$ as we can observe in Figs.\ref{fig:confin.pict},\ref{fig:ConfinJ2p}. On the other hand, if $J_4 \ne 0$ the bulk dimerization constrains the soliton to lie in the vicinity of the impurity (see Fig.\ref{fig:confin.pict}).

\paragraph{b) Enhancement of the magnetization near a dopant}

Under doping, the system becoming inhomogeneous, we define a local mean staggered magnetization
\begin{equation}
\label{MeStMg} {\mathcal{M}}^{\rm stag}_{i,a}=\frac{1}{4}
(-1)^{i+a}(2\langle S^{z}_{i,a} \rangle - \langle S^{z}_{i+1,a}
\rangle - \langle S^{z}_{i-1,a} \rangle)
\end{equation}
which has been calculated for a single dopant in a system of size $L\times M =16\times 8$. It is plotted for different values of the four-spin coupling in Fig. \ref{fig:ConfinJ2p} where the confinement mechanism can clearly be observed. Note that the inter-chain coupling induces a ``polarization cloud'' with strong AF correlations in the neighbor chains of the doped one.\\

\paragraph{c) Confinement length}

\begin{figure}
\centering
\psfrag{12}{$L=12$} \psfrag{L16}{$L=16$}
\psfrag{S}{{\small{$\langle S_{i,1}^z \rangle$}}}
\psfrag{z}{{\small{$i$}}} \psfrag{J4}{$J_{4}$} \psfrag{xi}{$\xi_{\parallel}$}
\psfrag{lo}{{\small{$2\xi_{\parallel}$}}} 
\includegraphics[height=6cm]{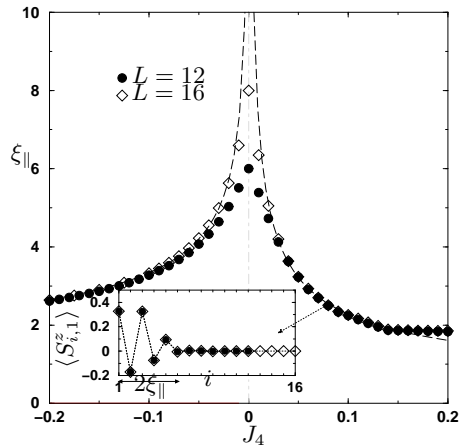}
\caption{ED data of the soliton average position  vs $J_4$
calculated for $\alpha=0.5$ and $J_{\bot}=0.1$. Different symbols
are used for $L\times M$ = $12\times 6$ and $16\times 8$ clusters.
The long-dashed line is a power-law fit (see text). Inset shows
the magnetization profile in the doped ($a=1$) chain at $J_4=0.08$, ie
$\xi_{\parallel} \simeq 2.5$ (Reprinted from Ref.\cite{ourprl03}).} 
\label{fig:xi}
\end{figure}

In order to measure the strength of confinement, a confinement length can be defined as
\begin{equation}
\label{xi.papp}
\xi_{\parallel}=\frac{\sum_{i}i|S_{i}^{z}|}{\sum_{i}|S_{i}^{z}|}.
\end{equation}
In the absence of confinement, the solitonic cloud is located at the center of the doped chain~:~$\xi_{\parallel}=L/2$. Otherwise, $\xi_{\parallel}$ converges to a finite value when $L \to \infty$. On Fig.\ref{fig:xi}, the confinement lenght is plotted versus $J_4$ for $2$ different system sizes at $\alpha=0.5$ and $J_{\perp}=0.1$. The finite size effects decrease for increasing $J_4$. Note that $\xi_{\parallel}(J_4)\neq\xi_{\parallel}(-J_4)$ and a
power law~\cite{Nakamura99} with different exponents $\eta$ is
expected when $J_4\rightarrow 0$. A fit gives $\eta\sim 0.33$ if
$J_4<0$ and $\eta\sim 0.50$ for $ J_4>0$ (Fig.\ref{fig:xi}). This
asymmetry can be understood from opposite renormalisations of
$J_1$ for different signs of $J_4$. Indeed, if $J_4<0$ then
$\delta J_{i,a}>0$ and the nearest neighbor MF exchange becomes
larger than the bare one. Opposite effects are induced by $J_4>0$.

\subsection{Effective interaction}
\begin{figure}
\centering
\psfrag{i}{$\Delta i$}
\psfrag{a}{(a) $\Delta a=0$}
\psfrag{b}{(b) $\Delta a=1$}
\psfrag{c}{(c) $\Delta a=2$}
\psfrag{d}{(d) $\Delta a=3$}
\psfrag{J}{{{$|J^{\rm eff}(\Delta a,\Delta
i)|$}}}
\includegraphics[height=6cm]{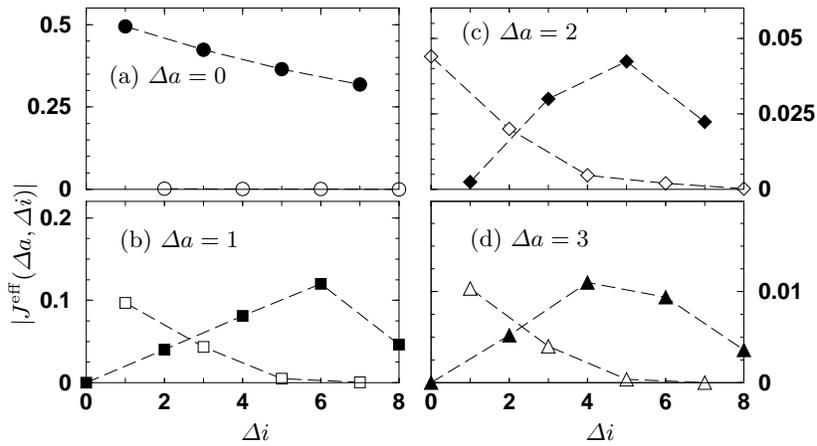} \caption{Magnitude of the
effective magnetic coupling between two impurities located either
on the same chain (a)  or on different ones (b-c-d)  vs the dopant separation $\Delta i$ in
a system of size $L \times M=16 \times 8$ with $\alpha=0.5$,
$J_{\bot}=0.1$, and $J_4=0.08$. Closed (resp. open) symbols
correspond to AF (F) interactions.}
\label{fig:Jeff}
\end{figure}

We now turn to the investigation of the effective interaction between dopants. A system of coupled chains with two dopants is considered here (see Fig.\ref{fig:lattice.cpled}). Each impurity releases an effective spin $\frac{1}{2}$, localized at a distance $\sim \xi_{\parallel}$ from it due to the confining potentiel set by $J_4$. 
 We
define an effective pairwise interaction $J^{\rm eff}$ as the
energy difference of the $S=1$ and the $S=0$ ground states. When $J^{\rm
eff}=E(S=1)-E(S=0)$ is positive (negative) the spin interaction is
AF (ferromagnetic). Let us first consider the case of two
dopants in the same chain. (i) When the two vacancies are on the
same sub-lattice the moments experience a very small ferromagnetic
$J^{\rm eff}<0$ as seen in Fig.~\ref{fig:Jeff} with $\Delta a=0$ so that the
two effective spins $\frac{1}{2}$ are almost free. (ii) When the
two vacancies sit on different sub-lattices, $\Delta i$ is odd and
the effective coupling is AF with a magnitude close to the
singlet-triplet gap. Fig.~\ref{fig:Jeff} with $\Delta a=0$ shows that the decay
of $J^{\rm eff}$ with distance is in fact very slow for such a
configuration. Physically, this result shows that a soliton and an
anti-soliton on the same chain and different sublattices tend to
recombine.

The behavior of the pairwise interaction of two dopants located on
{\it different} chains ($\Delta a=1,2,3,4$) is shown on
Fig.~\ref{fig:Jeff} for  $\Delta a=1,2,3,4$ for $J_4>0$. When dopants are on
opposite sub-lattices the effective interaction is
antiferromagnetic. At small dopant separation $J^{\rm eff}(\Delta
i)$ increases with the dopant separation as the overlap between
the two AF clouds increases until $\Delta i \sim 2\xi$. For larger
separation, $J^{\rm eff}(\Delta i)$ decays rapidly. Note that the
released spin-$\frac{1}{2}$ solitons bind on the opposite right
and left sides of the dopants as imposed by the the bulk
dimerisation~\cite{note2}. If dopants are on the same sub-lattice,
solitons are located on the same side of the dopants~\cite{note3}
and the effective exchange $J^{\rm eff}(\Delta i)$ is
ferromagnetic and decays rapidly to become negligible when $\Delta i >
2\xi$. The key feature here is the fact that the
effective pairwise interaction is {\it not} frustrating (because
of its sign alternation with distance) although frustration is
present in the microscopic underlying model. AF ordering is then
expected (at $T=0$) as seen for a related system of coupled
Spin-Peierls chains~\cite{Dobry98}.

\section{Conclusion}
The coexistence between AF order and SP dimer order under doping SP materials with non magnetic impurities~\cite{Zn_CuGeO} is one of the most surprising phenomenon in the field of quantum magnetism. Starting with the non frustating interaction between two solitonic clouds calculated above, we can construct an effective model of long range interacting spins $\frac{1}{2}$, randomly diluted on a square lattice. We have implemented a Quantum Monte Carlo (QMC) algorithm using the Stochastic Series Expansion (SSE) method~\cite{sse.sandvik} in order to study long distance interacting spin-$\frac{1}{2}$ models. The mechanism of AF ordering has been studied at very low temperature and dopant concentration with very large system sizes, up to $96 \times 96$~\cite{new.work}.
 
\label{sec:7}

\printindex
%
%

%

\begin{thebibliography}{99.}
\bibitem{note1} All operations of the point group will be defined
with respect to an (arbitrary) lattice site.
\bibitem{note2} In that case, the solution is unique and
the sign of the bulk dimerisation is fixed by the local condition
of maximum overlap between the 2 solitonic clouds.
\bibitem{note3} Two equivalent symmetric solutions are obtained with
opposite signs of the bulk dimerisation.
\end{thebibliography}
%


\end{document}